\documentclass[12pt]{iopart}
\usepackage{iopams}
\usepackage{graphicx}
\input amssym.def \input amssym

\begin{document}

\title[]{Ramanujan sums analysis of long-period sequences\\ and $1/f$ noise}

\author{Michel Planat,$^{\dag}$ Milan Minarovjech$^{\ddag}$ and Metod Saniga$^{\ddag}$}

\address{$^{\dag}$Institut FEMTO-ST, CNRS, 32 Avenue de
l'Observatoire,\\ F-25044 Besan\c con, France}

\address{$^{\ddag}$Astronomical Institute, Slovak Academy of Sciences,\\
SK-05960 Tatransk\'a Lomnica, Slovak Republic }

\begin{abstract}
Ramanujan sums 
are exponential sums with exponent defined over the irreducible fractions. 
Until now, they have been used to provide converging expansions to
some arithmetical functions appearing in the context of number
theory. In this paper, we provide an application of Ramanujan sum
expansions to periodic, quasi-periodic and complex time series, as
a vital alternative to the Fourier transform. The
Ramanujan-Fourier spectrum of the Dow Jones index over $13$ years
and of the coronal index of solar activity over $69$ years are
taken as illustrative examples. Distinct long periods may be
discriminated in place of the $1/f^{\alpha}$ spectra of the
Fourier transform.

\end{abstract}


\section{Introduction}

Signal processing of complex time-varying series is becoming more
and more fashionable in modern science and technology. Indices
arising from the stock market, changing global climate,
communication networks such as the Internet etc. are widely used
tools for business managers or governmental representatives. There
already exists a plethora of useful approaches for signal
processing of complex data. The oldest and perhaps most widely
used method is a Fourier analysis and its ``fast" implementation:
the fast Fourier transform (or FFT). Other complementary
techniques such as wavelet transforms, fractal analysis and
autoregressive moving average models (ARIMA) were developed with
the aim of identifying useful patterns and statistics in otherwise
seemingly random sequences \cite{Hamilton94}.

Ramanujan sums are defined as power sums over primitive roots of
unity. One can use an orthogonal property of these sums (closely
related to the orthogonal property of trigonometric sums) to form
convergent expansions of some arithmetical functions related to
prime number theory \cite{Rama18,Hardy21}.  Following the ideas of
Gadiyar and Padma \cite{Gadiyar99}, the first author proposed to
expand the domain of application of Ramanujan sum analysis from
number theory to arbitrary real time series and introduced the
concept of a {\it Ramanujan-Fourier transform}  \cite{Planat02}.
This earlier work remained quite ambiguous about the detection of
isolated periods. Ramanujan sum expansions of divisor sums, sums
of squares, and the Mangoldt function are well known.
Surprisingly, the detection of a singly periodic signal by the
Ramanujan sum analysis has not been considered before. But the
Ramanujan-Fourier amplitude corresponding to a single cosine
function of period $q$ is extremely simple: as we shall see, the
amplitude of the cosine function is simply scaled by the cosine of
the delay and the inverse of the Euler totient function
$\phi(n_0)$. Similarly to the standard discrete Fourier transform,
there are spurious signals of magnitude $O(n_0/t)$, depending of
the length $t$ of the averaging spectrum.

In the discrete Fourier transform, a sample to be analyzed is
discretized into pieces of length $1/q$ and the expansion is
performed over the $q$-th complex dimensional vectors of the
orthogonal basis $e_q^{(p)}(n):=\exp{(\frac{2i\pi p }{q}n)}$,
($p=1,\dots,q$). The orthogonal property reads $\sum_p
e_q^{(r)}(n)e_q^{(s)}(n)=q\delta(r,s)$, where $\delta(r,s)$ is the
Kronecker symbol. The expansion of a time series is $a(n)=\sum_p
a_pe_q^{(p)}(n)$ with Fourier coefficients $a_p=\frac{1}{q}\sum_n
a(n) e_q^{(p)}(-n)$, where the summation runs from $0$ to $q-1$.
In the Ramanujan-Fourier transform, the expansion $a(n)=\sum_q
a_qc_q(n)$ over the Ramanujan sums $c_q(n)=\sum'_pe_q^{(p)}(n)$
(see Sec.\,\ref{Rama}) involves the resolution  $\frac{1}{q}$ at
every single scale from $q=1$ to $t\rightarrow \infty$. The deep
principle behind rests on a very intricate link between the
properties of irreducible fractions $\frac{p}{q}$ and prime
numbers  \cite{Rama18}-\cite{Planat02}. As a result, one finds a
much finer structure of time series, with a variety of novel
features.

The paper is organized as follows. In Sec.\,\ref{Rama}, we remind
the reader with the arithmetical properties of Ramanujan sums,
provide the definition of the Ramanujan-Fourier transform and
examine the detection of a cosine signal. In Sec.\,\ref{Signals},
the use of the method is illustrated on the data from the stock
market and solar activity.

\section{Ramanujan sums and the Ramanujan-Fourier transform}
\label{Rama}

Ramanujan sums are real sums defined as $n$-th powers of $q$-th primitive roots of the unity,

$$c_q(n)=\sum_p' \exp{(2i\pi \frac{p}{q}n}),$$
where the summation runs through the $p$'s that are coprime to $q$
(hence the use of the symbol ``$'$"), being first introduced in the
context of number theory \cite{Rama18,Hardy21} for obtaining
convergent expansions of some arithmetical functions such as the
relative sum of divisors $\sigma(n)/n$ of an integer number $n$,

$$\sigma_{n}/n=\sum_{q=1}^{\infty} \frac{\pi^2}{6q^2} c_q(n).$$
They are multiplicative when considered as a function of $q$ for a fixed value of $n$, which can be used to prove an important relation

$$c_q(n)=\mu(q/q_1)\frac{\phi(q)}{\phi(q/q_1)},~q_1=(q,n).$$
In the above relation, the Euler totient function $\phi(q)$ is the
number of positive integers less than $q$ and coprime to it. The
M\"obius function, $\mu(n)$, vanishes if $q$ contains a square in
its (unique) prime number decomposition $\prod_i q_i^{\alpha_i}$
($q_i$ a prime number), and is equal to $(-1)^k$ if $q$ is the
product of $k$ distinct primes. One can readily checks the
following orthogonal property

$$\sum_{n=1}^{rs}c_r(n)c_s(n)=1~\mbox{if}~r=s ~~\mbox{and}~~\sum_{n=1}^q c_q^2(n)=q \phi(q) ~\mbox{otherwise}.$$
For an arithmetical function $a(n)$ possessing a Ramanujan-Fourier expansion

$$a(n)=\sum_{q=1}^{\infty} a_q c_q(n),$$
with Ramanujan-Fourier coefficients $a_q$, one can write a
Wiener-Khintchine formula,  relating the autocorrelation function
of $a(n)$ and its Ramanujan-Fourier power spectrum,

$$\mbox{lim}_{N \rightarrow \infty}\frac{1}{N}\sum_{n\le N} a(n)a(n+h)=\sum_{q=1}^{\infty}a_q^2 c_q(h).$$
This relation was used for counting the number of prime pairs
within a given interval \cite{Gadiyar99}. A similar formula has
been proposed for the convolution and cross-correlation
\cite{Wash08}.

Clearly, the Ramanujan sum analysis of an arithmetical function
looks like the Fourier signal processing of a time series $a(n)$
at discrete time intervals $n$. This formal analogy was developed
in \cite{Planat02} for the processing of time series with a rich
low frequency spectrum \cite{Planat01}. Ramanujan signal
processing was further developed in the context of quantum
information theory \cite{PlanatRosu03} eventually leading to an
original approach of quantum complementarity \cite{Planat05}. The
Ramanujan-Fourier transform was also used for processing time
series of the shear component of the wind at airports
\cite{Lagha06},  the structure of amino-acid sequences
\cite{Mainardi07} and in relation to the fast Fourier transform
\cite{Samadi05}. All these applications make use of the property
that for arithmetical functions possessing a mean value

$$A_v(x)=\mbox{lim}_{t \rightarrow \infty} \frac{1}{t} \sum_{n=1}^t a(n),$$
one can write the inversion formula

$$a_q=\frac{1}{\phi(q)} A_v(a(n) c_q(n)). $$

 \begin{figure}[ht]
\centerline{\includegraphics[width=7.0truecm,clip=]{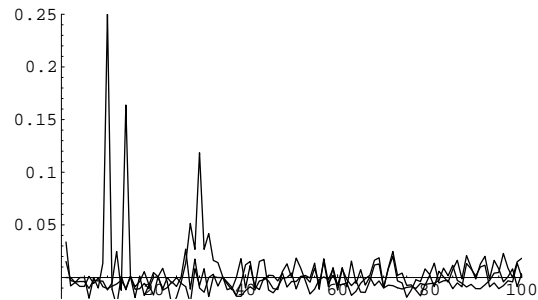}}
\caption{Ramanujan-Fourier spectrum for three different cosine
functions of periods $n_0=10$, $14$ and $30$, computed from a
sample of length $t=100$. The amplitude at $q=n_0$ equals
$1/\phi(n_0)= 1/4$, $1/6$ and $1/8$, respectively. }
\end{figure}
\begin{figure}[ht]
\centerline{\includegraphics[width=7.0truecm,clip=]{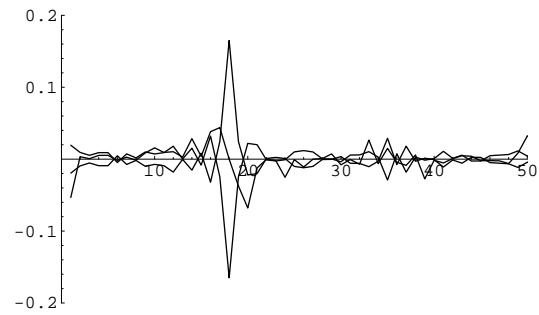}}
\caption{Ramanujan-Fourier spectrum for the cosine function of
period $n_0=38$,  with delays $\delta=0$, $\pi/2$ and $\pi$, and
sample length $t=100$. The amplitudes of the peaks for $\delta=0$
and $\pi$ is $1/6$.  }
\end{figure}
\begin{figure}[ht]
\centerline{\includegraphics[width=7.0truecm,clip=]{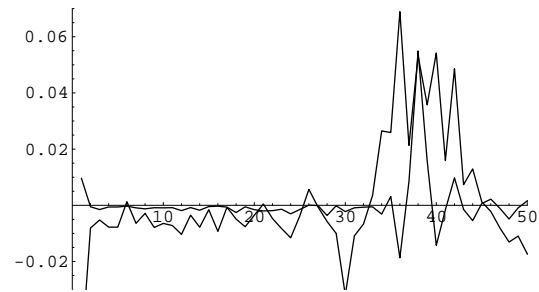}}
\caption{Ramanujan-Fourier spectrum for the cosine function of
period $n_0=18$,  with sample lengths $t=100$ and $500$. One
clearly observes the compression of the lines when $t$ increases.
}
\end{figure}

\subsection*{Ramanujan-Fourier transform of a cosine function}

Let us consider now the Ramanujan signal processing of a periodic (cosine) function of period $n_0$

$$a(n)=a_0 \cos (2 \pi \frac{n}{n_0} +\delta).$$
The Ramanujan-Fourier coefficients read

$$a_q=\mbox{lim}_{t \rightarrow \infty}\frac{a_0 \exp(i\delta)}{2\phi(q)t}\sum'_p\sum_{n=1}^t \exp[2 i \pi n(\frac{p}{q}+\frac{1}{n_0})]+c.c.$$
in which the order of summations is reversed and ``c.c." stands
for the complex conjugate. Assume first that $t$ is a multiple of
the period $n_0$. Then the $n$-th summation is zero unless
$k=\frac{p}{q}+\frac{1}{n_0}$ is a positive integer. For instance,
$a_q$ can be non-zero if $q=n_0$ under the conditions that $n_0$
divides $p+1$ and $k=1$, i.\,e. $p=n_0-1$ (otherwise $p>q$, which is
outside the range of summation of the $p$ sum). One then gets

$$a_{n_0}=\mbox{lim}_{t \rightarrow \infty}\frac{a_0 \exp (i\delta)}{2\phi(n_0)t}\sum_{n=1}^t\exp (2i\pi\frac{ n}{n_0})+c.c.$$
The $n$-th summation equals zero unless $n_0$ divides $n$,
otherwise it equals to $t$. As a result, the amplitude of the
$n_0$-th line reflects the amplitude of the periodic signal as

$$a_{n_0}=\frac{a_0}{\phi(n_0)}\cos (\delta).$$
In general, $t$ is not a multiple of $n_0$ so that there exists an
extra contribution to the amplitude, of order of magnitude
$O(\frac{n_0}{t})$. As long as the period $n_0$ is much smaller than
the length $t$ of the sample, i.\,e. $n_0\ll t$, one observes a
single line at $n_0$; otherwise bursts of non-zero amplitudes
emerge in the vicinity of the lines $k n_0$ --- see Figs.\,1--3.

Thus, there are two significant differences when compared to a
period analysis by the standard discrete Fourier transform. First,
the amplitude of the line at the period $n_0$ is scaled by a
factor of $\phi(n_0)$. Second, the Ramanujan-Fourier analysis is
sensitive to the delay $\delta$. The latter feature may, at first
sight, seem as a drawback since some period of the signal to be
analyzed may be hidden by the dephasing effect. One method to
circumvent this difficulty is to average the spectra corresponding
to several shifted samples of the signal.

\subsection*{Ramanujan-Fourier transform of a period modulated cosine function}

Let us now apply the approach to a period modulated cosine
function. We intentionally select a period modulation with a large
index (equal to $1$). The selected modulation is

$$n_0=n_0[1+ \sin(2 \pi \frac{n}{n_1}) ],$$
with $n_0=10$ and $n_1=14$. The sample length is $t=2000$. Due to
a high modulation index, the FFT analysis (shown in Fig.\,4) does
not easily allow to recover the constituent integer periods $10$
and $14$. In contrast, the Ramanujan sum analysis is very powerful
in this context. From Fig.\,5 one clearly identifies (positive)
large amplitudes at the periods $10$, $12$, $2\times 12$ and
$\mbox{LCM}(10,12)=70$ (LCM being the least common multiple). Thus,
for an input signal of the period $n_0$ and period modulation
$n_1$, the FFT exhausts all lines at $l n_0 + m n_1$ ($l$ and $m$
integers), eventually leading to a continuous spectrum in the
limit of incommensurate periods $n_0$ and $n_1$. In contrast, the
Ramanujan-Fourier transform is straightforward in identifying the
input modulation.

\begin{figure}[ht]
\centerline{\includegraphics[width=7.0truecm,clip=]{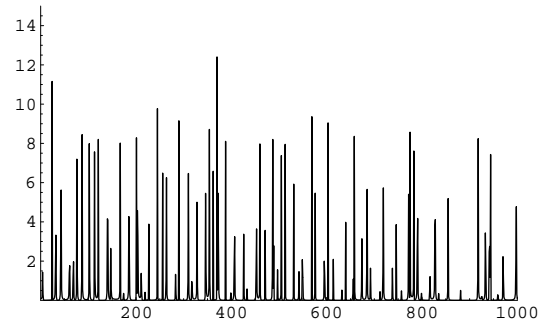}}
\caption{FFT spectrum of a period modulated cosine: $n_0=10$ and $n_1=14$. }
\end{figure}
\begin{figure}[ht]
\centerline{\includegraphics[width=7.0truecm,clip=]{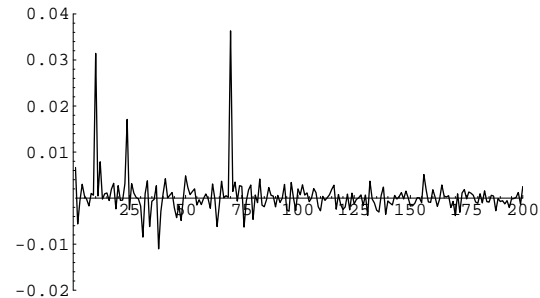}}
\caption{Ramanujan-Fourier transform of a period modulated cosine function: $n_0=10$ and $n_1=14$.}
\end{figure}

\section{Ramanujan sums analysis of some complex systems}
\label{Signals}

As a nice illustration of the above-outlined properties the
Ramanujan sum analysis, we shall analyze a couple of complex time
series taken form the stock market and solar activity.

\subsection*{The Dow Jones index of the stock market}

The first time series deals with the Dow Jones index and has been downloaded from\\
http : // www.optiontradingtips.com/resources/historical - data/dow-jones30.html.
\\
Fig.\,6 depicts the evolution of Dow Jones 30 Industrials stock
price over about $13$ years. The power spectral density of the
prices (Fig.\,8) approximately follows a $1/f^2$ law versus the
Fourier frequency $f=n^{-1}$, compatible with a
Brownian-motion-based model \cite{Osborne59,Bouchaud2000}. The
Ramanujan-Fourier analysis shown in Fig.\,7 yields a more detailed
structure with many (positive or negative) peaks centered at
well-identified frequencies. There exists a sensitivity of the
amplitude of the peaks (not shown) on the number $t$ of data, but
the position of the peaks, as well as their statistics, is not
dependent on $t$. Since both spectra in Figs. 6 and 7 are given in
a logarithmic time-scale, it follows that the Ramanujan sum
analysis provides a clear advantage over the standard Fourier
analysis in offering a rich and structured signature. Here, we
shall not delve any further into the origin of this structure,
which will be a topic of a separate paper.

\begin{figure}[ht]
\centerline{\includegraphics[width=7.0truecm,clip=]{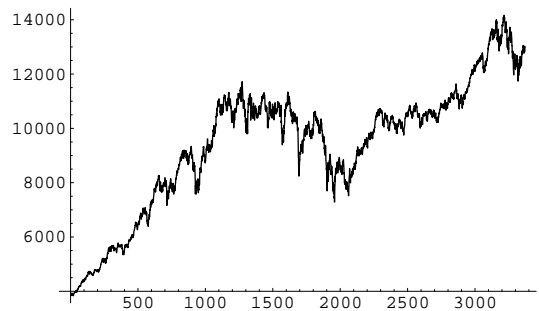}}
\caption{The Dow Jones 30 Industrials from 03.01.1995 to 30.05.2008.}
\end{figure}
\begin{figure}[ht]
\centerline{\includegraphics[width=7.0truecm,clip=]{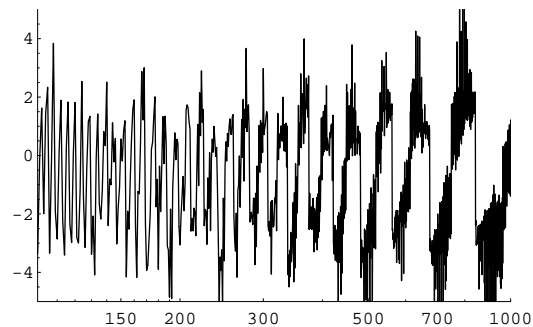}}
\caption{The Ramanujan sums analysis of the Dow Jones 30
Industrials. Periods in the range $n=100$ to $n=1000$ were
selected. One clearly sees (positive and negative) peaks centered
about non-equally-spaced periods. }
\end{figure}
\begin{figure}[ht]
\centerline{\includegraphics[width=7.0truecm,clip=]{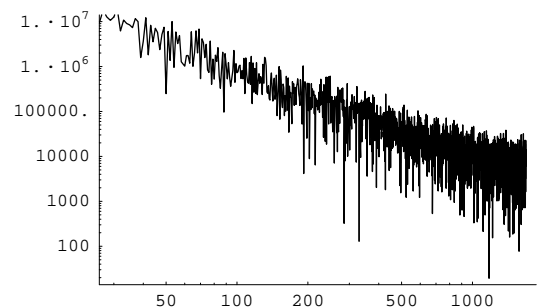}}
\caption{The FFT analysis of the Dow Jones 30 Industrials in a
log-log scale. The standard $1/f^2$ dependence of the power
spectrum, which is characteristic of a Brownian-like motion, is
clearly visible. }
\end{figure}

\subsection*{The coronal index of the solar activity}

The second time series has been picked up from\\
http : // www.ngdc.noaa.gov/stp/SOLAR/ftpsolarcorona.html\#index.
\\
It represents the Green Line (FeXIV 530.3 nm) Coronal Index of
solar activity from 1939 to 2008. One easily recognizes from Figs.
9 and 10 that the coronal index is approximately periodic, with a
period about 10 years. The whole FFT spectrum shown in Fig 11
exhibits a $1/f$ dependence characteristic of many  physical,
biological, arithmetical \cite{Planat01, Planat02, Li1996} and
other complex systems \cite{Li1996}.

\begin{figure}[ht]
\centerline{\includegraphics[width=11.0truecm,clip=]{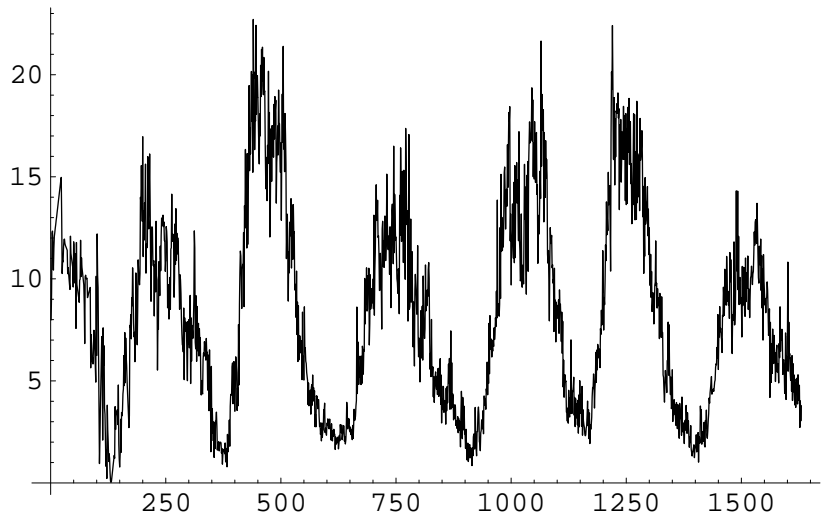}}
\caption{The temporal variation of the Coronal Index; a $10$-year period is clearly visible.}
\end{figure}
\begin{figure}[ht]
\centerline{\includegraphics[width=11.0truecm,clip=]{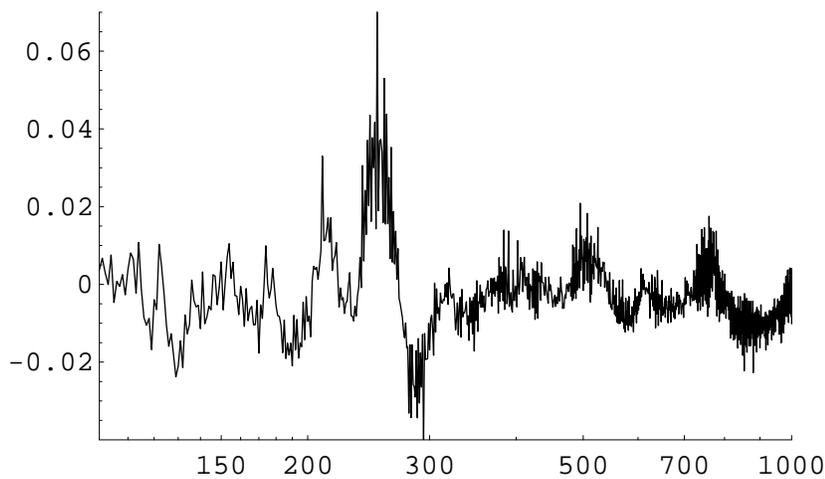}}
\caption{The RFT of the Coronal Index. The 10-year period is
clearly identified; other longer periods of a smaller amplitude
are observed as well, besides the harmonics $pn_0$ ($p$ integer).}
\end{figure}
\begin{figure}[ht]
\centerline{\includegraphics[width=11.0truecm,clip=]{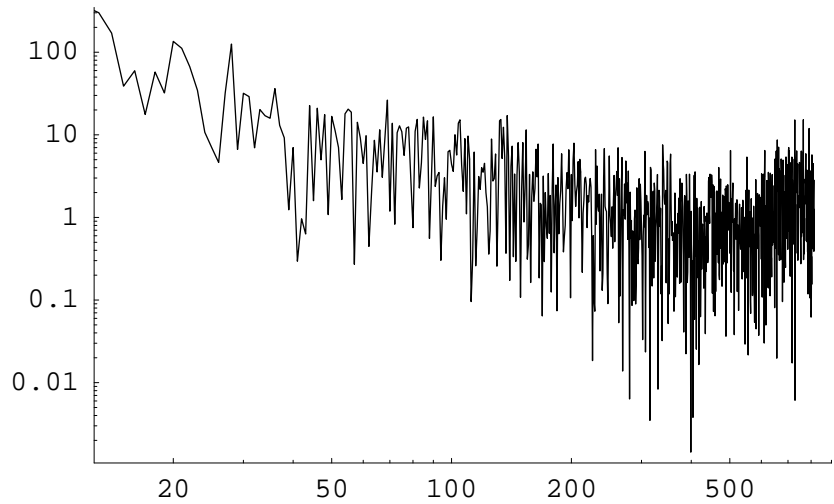}}
\caption{The FFT of the Coronal Index. One observes a $1/f$
dependence of the power spectrum; the $10$-year period is quite
hard to identify.}
\end{figure}

\section*{Perspectives and Conclusion}

It is a widely shared belief that $1/f^{\alpha}$ noises are so
random that non-statistical models of them are currently out of
reach. A counterexample to this belief can be found in
\cite{Planat01}, in which an arithmetical approach to $1/f$ noise
was suggested. In the present paper, we offer another perspective
by analyzing the data from an arithmetical magnifying glass built
on Ramanujan sums. The Ramanujan-Fourier transform is able to
extract quasi-periodic features which are characteristic of number
theoretical functions \cite{Rama18}-\cite{Planat02}, as well as
fine periodic features that the standard Fourier transform may
hide. A Ramanujan sums analysis is a multi-scale prism with scales
related to each other by the properties of irreducible fractions.
It is particularly well-suited for analyzing rich time series
showing a $1/f^\alpha$ ($0<\alpha< 2$) FFT dependence. We selected
two specific complex systems to illustrate the power of this new
method: the data from the stock market (for which the price index
FFT follows a $1/f^2$-law) and those from solar cycle activity
(for which the coronal index follows a $1/f$-law). A more detailed
examination of the latter will be given in a separate paper.

\end{document}